\documentclass[a4paper,11pt]{article}
\usepackage{jheppub}
\usepackage[T1]{fontenc}
\usepackage{mathtools}
\usepackage{enumitem}
\hypersetup{pdftitle={The type-II double copy for relativistic Rindler fluids},
  pdfsubject={Construction in a unified framework, analysis of the degeneration of solution families, and comparison of leading-order nonrelativistic limits}}
\newcommand{\cB}{\mathcal B}
\newcommand{\cK}{\mathcal K}
\newcommand{\rt}{\tilde r}
\newcommand{\dd}{\mathrm d}
\newcommand{\ii}{\mathrm i}
\newcommand{\ord}[1]{\mathcal O(#1)}
\newcommand{\RePart}{\operatorname{Re}}
\newcommand{\ImPart}{\operatorname{Im}}

\newcommand{\KeelerMongaEq}[1]{\href{https://arxiv.org/pdf/2404.03195v1\#nameddest=equation.#1}{(#1)}}
\newcommand{\WangEtAlEq}[1]{\href{https://arxiv.org/pdf/2609.29922v1\#nameddest=equation.#1}{(#1)}}

\title{The type-II double copy for relativistic Rindler fluids}

\author{Yu-Cheng Wang$^{1,3,4}$}
\author{Jing-Rui Zhang$^{1,3,4}$}
\author{Yun-Long Zhang$^{2,1}$}
\affiliation{$^{1}$ School of Fundamental Physics and Mathematical Sciences,  Hangzhou   Institute for Advanced Study, UCAS, Hangzhou 310024, China. }
\affiliation{$^{2}$ National Astronomical Observatories, Chinese Academy of Sciences, Beijing, 100101, China}
\affiliation{$^{3}$CAS Key Laboratory of Theoretical Physics, Institute of Theoretical Physics, Chinese Academy of Sciences, Beijing 100190, China.} 
\affiliation{$^{4}$Taiji Laboratory for Gravitational Wave Universe(Beijing/Hangzhou), University of Chinese Academy of Sciences, Beijing 100149, China.}

\emailAdd{wangyucheng25@mails.ucas.ac.cn,zhangjingrui22@mails.ucas.ac.cn,\\
zhangyunlong@nao.cas.cn}

\abstract{Within the Weyl double copy framework, we study a mixed single-copy construction for the Petrov type-II geometry of first-order relativistic Rindler fluids. Working on a common fixed flat Rindler background with consistent spinor conventions and a real field strength representation, we algebraically factorize the Weyl spinor into two symmetric rank-two Maxwell spinors and test the Maxwell equations and the zeroth copy scalar equation up to the first order in the relativistic gradient expansion. We separate the degeneration parameter from the gradient expansion parameter and analyze the continuous degeneration to the Petrov type-D and type-N branches as the complex shear and vorticity tend to zero, together with the preservation of the accuracy of the Maxwell and scalar field equations along each path. We also compare the relativistic construction with the nonrelativistic case.
}

\begin{document}
\maketitle
\flushbottom

\section{Introduction}
\label{sec:intro}

The AdS/CFT correspondence relates gravitational dynamics in asymptotically anti-de Sitter spacetimes to observables of a boundary conformal field theory~\cite{Maldacena1998,Gubser1998,Witten1998}. Early holographic studies connected thermal response and transport properties to the dynamics of black branes and horizon data~\cite{Policastro2001,SonStarinets2002,Kovtun2005,IqbalLiu2009}.
Within this broader context, the fluid/gravity correspondence provides a systematic framework for exploring the relation between fluid dynamics and spacetime geometry order by order~\cite{Bhattacharyya2008,Rangamani2009,Hubeny2012,Bredberg2011}. In the Rindler fluid/gravity construction, the fluid pressure and velocity on a finite timelike cutoff hypersurface are promoted to slowly varying boundary fields, whose stress tensor admits a quasilocal gravitational interpretation~\cite{BrownYork1993}. The bulk vacuum Einstein equations can then be solved order by order in a long-wavelength expansion, with the gravitational constraints reducing to the fluid equations on the cutoff surface~\cite{Compere2011,Bredberg2012}. Related derivations use horizon membrane dynamics~\cite{ElingFouxonOz2009,ElingOz2010} or impose a Petrov type-I condition on a timelike hypersurface~\cite{LysovStrominger2011,Huang2011}. Near-horizon deformations of Schwarzschild spacetime yield an incompressible fluid on the sphere~\cite{BredbergStrominger2012}. The relativistic Rindler construction uses a covariant gradient expansion~\cite{Compere2012,Eling2012}. Further work examines the horizon entropy current~\cite{ElingEntropy2012}, constraints on second-order transport~\cite{MeyerOz2013}, and the relation between the Petrov type-I condition and the dual fluid dynamics~\cite{CaiLiYangZhang2013,Cai2014}.

A complementary link between gravity and gauge theory is provided by the double copy. At the level of scattering amplitudes, the double copy programme originates in the Kawai-Lewellen-Tye relations between closed- and open-string tree amplitudes~\cite{Kawai1986} and in color-kinematics duality~\cite{Bern2008,Bern2010}. Comprehensive reviews can be found in refs.~\cite{BernReview2024,BernSAGEX2022}. The classical double copy extends this paradigm to exact and perturbative field solutions, establishing a constructive map between gauge and gravitational fields~\cite{Monteiro2014,Luna2019,Godazgar2021,White2021,Chacon2021,Adamo2022Snowmass}. Key developments include Taub-NUT and static Kerr-Schild constructions~\cite{LunaTaubNUT2015,RidgwayWise2016}, time-dependent solutions and radiation from color charges~\cite{LunaRadiation2016,GoldbergerRidgway2017}, and nonperturbative solutions of biadjoint scalar theory~\cite{WhiteScalar2016}. Related geometric formulations include the Newman-Penrose map from a class of exact gravitational solutions to self-dual Maxwell fields~\cite{Elor2020NPMap} and Weyl doubling constructions with gauge fields on curved gravitational backgrounds~\cite{Alawadhi2020WeylDoubling}. Twistor methods provide alternative formulations~\cite{Chacon2022Alternative} and clarify the locality of the spacetime relation for selected type-D solutions~\cite{Luna2022Locality}. Further extensions encompass asymptotic Weyl double copies~\cite{Godazgar2021Asymptotic} as well as sourced formulations for type-D and Einstein-Maxwell geometries~\cite{Easson2021Sources,Easson2023EinsteinMaxwell}, with recent spinorial derivations applying generally at the linearized level and admitting exact extensions in special cases~\cite{ArmstrongWilliams2025Sources}.

For the four-dimensional type-D and type-N solutions studied in Refs.~\cite{Luna2019,Godazgar2021}, the Weyl double copy expresses the Weyl curvature spinor as the symmetrized square of a Maxwell spinor divided by a complex zeroth copy scalar. A Petrov type-II Weyl spinor requires two distinct Maxwell factors~\cite{Keeler2024}. The connection to nonrelativistic Rindler fluids~\cite{Keeler2020} was extended to type-II bulk metrics in Ref.~\cite{Keeler2024}, which analyzes the algebraic factorizations and the Maxwell and scalar field equations in a near-horizon expansion.
Complementing these nonrelativistic constructions, studies of first-order relativistic Rindler fluids show that, once the first-order inviscid fluid constraint equations are imposed, the leading-order bulk Weyl curvature is determined entirely by the fluid vorticity and shear~\cite{RelDraft}. At first order in the gradient expansion, the bulk geometry for a general fluid flow is of Petrov type II. It degenerates to Petrov type D when the shear vanishes at nonzero vorticity, and to Petrov type N when the vorticity vanishes at nonzero shear. Reference~\cite{RelDraft} systematically derives the single-copy fields of the relativistic type-D and type-N branches together with their nonrelativistic limits, explicitly distinguishing exact source-free solutions with a constant magnetic field from local perturbative solutions with slow variation at first order in gradients.

These results motivate three questions for the first-order relativistic type-II double copy:
(1) Can the fields obtained from the known algebraic factorizations satisfy the Maxwell and scalar equations on a common fixed flat Rindler background at the prescribed gradient orders?
(2) Can continuous degeneration recover the relativistic type-D/N single copies while preserving the required residual orders at finite values of the degeneration parameter?
(3) How should coordinate scaling, null-frame normalization, and spinor chirality conventions be transformed to compare the leading coefficients of the same real electromagnetic field strength in the nonrelativistic limit?

In this paper, we address these questions using a common $P,Q,R$ spinor expansion and real field strength representation on a fixed flat Rindler background. We reconstruct the real electromagnetic field strengths from the complex spinor coefficients. Comparisons are then performed at the level of \(F_{\mu\nu}\), with the frame, normalization, and chirality conventions consistently aligned. On the gravitational side, we retain the bulk Weyl curvature through first order in gradients. Under the chosen gradient assignments and local nondegeneracy conditions, branch A satisfies the required Maxwell and scalar residual estimates. Separating the degeneration parameter from the gradient expansion parameter, we establish uniform residual estimates along the type-D and $N_1$ paths and identify the additional differential conditions required for the fixed-amplitude $N_0$ construction. Finally, we derive the transformations to the nonrelativistic frame and compare leading-order double copy structures within amplitude subfamilies obeying the same conditions on functional dependence.

The paper is organized as follows.
Section~\ref{ch:construction} constructs the relativistic type-II double copy, tests the two algebraic branches, and presents the symmetric magnetic normalization. Section~\ref{ch:degeneration-recalculated} analyzes the type-D/N degenerations and their residual estimates. Section~\ref{b:chapter} derives the nonrelativistic limit of the type-II double copy and compares the leading-order structures. Section~\ref{ch:summary} summarizes the results and open questions.

\section{Type-II double copy for first-order relativistic Rindler fluids}

\label{ch:construction}

In this section, we construct a first-order Petrov type-II double copy on a fixed flat Rindler background, unifying the spinor and real field strength representations. Order-by-order checks show that branch A satisfies the required perturbative accuracy, whereas branch B is excluded under the prescribed order assignment. Adopting a symmetric magnetic normalization further reveals a common magnetic component accompanied by radiative components of equal magnitude and opposite sign in the two single-copy fields, which naturally reduces to the purely magnetic type-D configuration in the shear-free limit.

\subsection{Rindler fluid geometry and Weyl scalars}
\label{a:setting}

The Weyl double copy establishes a correspondence between four-dimensional gravitational and gauge fields at the level of curvature. Its basic form is~\cite{Luna2019}
\begin{equation}
 C_{ABCD}=\frac{1}{S}f_{(AB}f_{CD)},
 \label{a:weyl-double-copy}
\end{equation}
where $C_{ABCD}$ is the Weyl curvature spinor on the gravitational side, $f_{AB}=f_{BA}$ is the Maxwell spinor representing the electromagnetic field strength, and $S$ is the complex scalar, which is regarded as the zeroth copy. This relation is defined in a local region where $S\ne0$, and parentheses $f_{(AB}f_{CD)}$ denote complete symmetrization of the four spinor indices with normalized weight. For the general Petrov type-II case studied here, the two Maxwell spinor factors need to be allowed to differ, leading to the mixed Weyl double copy form~\cite{Keeler2024}. Its explicit expression and algebraic decomposition are presented in subsection~\ref{a:dictionary}. In addition to satisfying this algebraic reconstruction relation, the single-copy fields and the zeroth copy scalar must obey the source-free Maxwell equations and the scalar wave equation, respectively, on the fixed flat Rindler background specified below. In our perturbative construction, these field equations are verified to the prescribed order.

Specifically, we adopt the two algebraic factorizations of the type-II Weyl spinor presented in the nonrelativistic type-II double copy study~\cite{Keeler2024} and test the Maxwell and scalar field equations at the corresponding order in the relativistic gradient expansion. On the gravitational side, we retain the bulk Weyl curvature through first order in gradients. On the gauge theory side, we adopt the assignments of gradient orders $f^{(i)}\sim\ord{\partial}$ and $S\sim\ord{\partial}$, requiring the zeroth- and first-order terms in the exterior derivative $\dd F^{(i)}$ and divergence $\bar\nabla_\mu F^{(i)\mu\nu}$ to vanish identically while allowing higher-order residuals of $\ord{\partial^2}$. The zeroth copy scalar equation is likewise checked according to its assigned gradient order.

Let the timelike cutoff hypersurface be located at $r=1$, with induced boundary metric $\eta_{ab}=\operatorname{diag}(-1,1,1)$ and normalized fluid velocity $u^au_a=-1$.
We define the transverse projector \mbox{$h_{ab}=\eta_{ab}+u_au_b$}, the convective derivative \mbox{$D=u^a\partial_a$}, and the fluid acceleration \mbox{$a_a=Du_a$}. The kinematic quantities are the expansion tensor \mbox{$K_{ab}=h_a^ch_b^d\partial_{(c}u_{d)}$}, the vorticity tensor \mbox{$\Omega_{ab}=h_a^ch_b^d\partial_{[c}u_{d]}$}, and the expansion scalar \mbox{$\Theta=\partial_au^a$}. Here the index symmetrization and antisymmetrization both include a factor of \mbox{$1/2$}. In the relativistic gradient expansion, the fluid pressure $p$ and velocity $u^a$ are of order $\ord{1}$. Each boundary derivative $\partial_a$ raises the gradient order by one, whereas the radial derivative $\partial_r$ leaves the gradient order unchanged. We take the bulk metric and first-order fluid constraint equations directly from the first-order relativistic Rindler fluid single-copy study~\cite{RelDraft} (see Ref.~\cite{Compere2012} for details of the underlying fluid/gravity construction):
\begin{equation}
 \begin{aligned}
 \dd s^2={}&-2pu_a\dd x^a\dd r+
 \Big\{\eta_{ab}-p^2(r-1)u_au_b\\
 &\quad+2(r-1)[u_au_bDp+2pa_{(a}u_{b)}]\Big\}\dd x^a\dd x^b
 +\ord{\partial^2},
 \end{aligned}
 \label{a:metric}
\end{equation}
\begin{equation}
 \Theta\sim\ord{\partial^2},\qquad
 a_a+h_a^b\partial_b\ln p\sim\ord{\partial^2}.
 \label{a:metric-euler}
\end{equation}

Classical double copies have also been formulated on curved and maximally symmetric backgrounds~\cite{BahjatAbbas2017,CarrilloGonzalez2018}, and related spinorial constructions connect Weyl curvature to massless free fields in curved spacetimes~\cite{Han2022FreeFields} or treat selected vacuum solutions with a cosmological constant~\cite{Han2022Cosmological}. In the present work, we define the background metric for the single-copy fields by holding the zeroth-order fluid parameters fixed, following the convention of Ref.~\cite{RelDraft}.
Specifically, a boundary Lorentz transformation with constant parameters is used to align with the local rest frame at a chosen reference boundary point, and the background pressure is frozen to its positive constant value $p>0$ at that point. The Maxwell and scalar field equations are therefore formulated on the flat Rindler background metric
\begin{equation}
 \dd\bar s^2=-\rt \dd\tau^2+2p \dd\tau\dd r+
 \dd x^2+\dd y^2,\qquad \rt=1+p^2(r-1),
 \label{a:background}
\end{equation}
in the domain $\rt>0$. This background is exactly flat, and the frozen background parameter $p$ is treated as a constant under boundary differentiation.

In a comoving orthonormal spatial frame, we define the vorticity component, the vorticity-dependent amplitude, and the complex shear component by
\begin{equation}
 \Omega=\Omega_{12},\qquad \cB=p\Omega,\qquad
 \cK=K_{11}-\ii K_{12},\qquad K_{11}+K_{22}\sim\ord{\partial^2}.
\label{a:vorticity-shear}
\end{equation}
Upon imposing the first-order fluid equations of motion~\eqref{a:metric-euler}, Eq.~\WangEtAlEq{37} of Ref.~\cite{RelDraft} yields the nonvanishing first-order Weyl scalars
\begin{equation}
 \Psi_2=\frac{\ii p\Omega}{2},\qquad
 \Psi_4=-\frac{p\cK}{\rt},\qquad
 \Psi_0=\Psi_1=\Psi_3=0.
 \label{a:weyl}
\end{equation}
According to the Petrov classification criteria summarized in table~\href{https://arxiv.org/pdf/2609.29922v1\#nameddest=table.1}{I} of Ref.~\cite{RelDraft}, the first-order bulk geometry is of Petrov type II when $\Omega\cK\ne0$. It degenerates to Petrov type D when $\cK=0$ and $\Omega\ne0$, and to Petrov type N when $\Omega=0$ and $\cK\ne0$. We emphasize that this algebraic classification characterizes the Weyl spinor truncated at first order in the gradient expansion and does not by itself determine the exact Petrov type of the full metric once higher-order gradient corrections are included.

\subsection{Spinor conventions and algebraic decomposition }
\label{a:dictionary}

We adopt the null coframe of Eq.~\WangEtAlEq{27} of Ref.~\cite{RelDraft},
\begin{equation}
\begin{aligned}
 \ell&=-\sqrt{\rt/2} \dd\tau,\qquad
 &n&=-\sqrt{\rt/2} \dd\tau+p\sqrt{2/\rt} \dd r,\\
 m&=\frac{-\ii\dd x+\dd y}{\sqrt2}, &\bar m&=\frac{\mathrm i\mathrm dx+\mathrm dy}{\sqrt2}
 \label{a:tetrad}
\end{aligned}
\end{equation}
in terms of which the fixed Rindler background metric takes the form $\bar g=-2\ell_{(\mu}n_{\nu)}+2m_{(\mu}\bar m_{\nu)}$. We expand a general symmetric rank-two Maxwell spinor as
\begin{equation}
 f_{AB}=P o_{(A}\iota_{B)}+Q o_Ao_B+R \iota_A\iota_B.
 \label{a:selfdual-map}
\end{equation}
With the standard spin dyad normalization
\begin{equation}
o_A\iota^A=1,\qquad
\iota_Ao^A=-1,\qquad
o_Ao^A=\iota_A\iota^A=0,
\label{a:dyad-normalization}
\end{equation}
the three Maxwell NP scalars are related to the spinor expansion coefficients by
\begin{equation}
 \phi_0=F(\ell,m)=R,\qquad
 \phi_1=\frac12\{F(n,\ell)+F(m,\bar m)\}=-\frac P2,
 \qquad \phi_2=F(\bar m,n)=Q.
 \label{a:correct-phi}
\end{equation}
The mapping from the spinor coefficients to the physical real electromagnetic field strength is uniquely fixed by the null coframe, spacetime orientation, and NP conventions above, with the complex two-form normalized as
\begin{equation}
 \mathcal F=P(m\wedge\bar m-\ell\wedge n)
 +2Q \ell\wedge m-2R n\wedge\bar m,\qquad F=\RePart\mathcal F.
\label{a:complex-F}
\end{equation}
Expanded in the coordinates $(\tau,r,x,y)$, the real field strength two-form reads explicitly
\begin{equation}
 \begin{aligned}
 F={}&p\RePart P \dd\tau\wedge\dd r
 +\ImPart P \dd x\wedge\dd y\\
 &-\sqrt{\rt} \dd\tau\wedge
 \left[(\ImPart Q+\ImPart R) \dd x
 +(\RePart Q-\RePart R) \dd y\right]\\
 &+\frac{2p}{\sqrt{\rt}} \dd r\wedge
 \left[\ImPart R \dd x-\RePart R \dd y\right].
 \end{aligned}
 \label{a:full-F}
\end{equation}

For the four-dimensional Weyl spinor, we use the fully symmetrized expansion of Eq.~\WangEtAlEq{38} of Ref.~\cite{RelDraft}:
\begin{equation}
 \begin{aligned}
 C_{ABCD}={}&\Psi_0\iota_A\iota_B\iota_C\iota_D
 -4\Psi_1o_{(A}\iota_B\iota_C\iota_{D)}
 +6\Psi_2o_{(A}o_B\iota_C\iota_{D)}\\
 &-4\Psi_3o_{(A}o_Bo_C\iota_{D)}+\Psi_4o_Ao_Bo_Co_D.
 \end{aligned}
 \label{a:weyl-spinor}
\end{equation}
For two general Maxwell spinors $f^{(i)}_{AB}=P^{(i)} o_{(A}\iota_{B)}+Q^{(i)} o_Ao_B+R^{(i)}\iota_A\iota_B$ ($i=1,2$), the algebraic mixed Weyl double copy relation $C_{ABCD}=S^{-1}f^{(1)}_{(AB}f^{(2)}_{CD)}$ is equivalent to the five algebraic equations
\begin{equation}
 \begin{aligned}
 R^{(1)}R^{(2)}&=S\Psi_0,& P^{(1)}R^{(2)}+P^{(2)}R^{(1)}&=-4S\Psi_1,\\
 P^{(1)}P^{(2)}+Q^{(1)}R^{(2)}+Q^{(2)}R^{(1)}&=6S\Psi_2,&
 P^{(1)}Q^{(2)}+P^{(2)}Q^{(1)}&=-4S\Psi_3,\\
 Q^{(1)}Q^{(2)}&=S\Psi_4.
 \end{aligned}
 \label{a:five-algebraic}
\end{equation}

For a Petrov type-II Weyl spinor with $\Psi_0=\Psi_1=\Psi_3=0$ and $S\Psi_2\Psi_4\ne0$, the solutions of Eq.~\eqref{a:five-algebraic} have the two branch structures identified in Eqs.~\KeelerMongaEq{4.46} and~\KeelerMongaEq{4.47} of Ref.~\cite{Keeler2024}. The local nondegeneracy condition requires that $S\Psi_2\Psi_4\neq 0$ at every point of the local patch under consideration. Up to factor exchange, there are two inequivalent branches:
\begin{itemize}
 \item Branch A: $R^{(1)}=R^{(2)}=0$, 
 \item Branch B: $P^{(1)}=P^{(2)}=R^{(1)}=0$ and $R^{(2)}\ne0$.
\end{itemize}

Introducing the radially rescaled complex amplitudes
\begin{equation}
 H=\sqrt{\rt} Q,\qquad L=\frac R{\sqrt{\rt}},
 \label{a:PHL}
\end{equation}
where the physical real field strength two-form continues to be determined by the real and imaginary parts of $P,Q,R$ via Eq.~\eqref{a:full-F}, we can express branch A as
\begin{equation}
 \begin{gathered}
 P^{(1)}=\xi H^{(1)},\qquad P^{(2)}=-\xi H^{(2)},\qquad Q^{(i)}=H^{(i)} /\sqrt{\rt},
 \\
 R^{(1)}=R^{(2)}=0, \qquad
 S=-\frac{H^{(1)}H^{(2)}}{p\cK},
 \end{gathered}
 \label{a:branch-A}
\end{equation}
and branch B as
\begin{equation}
 \begin{gathered}
 P^{(1)} =P^{(2)}=R^{(1)}=0,\qquad
 Q^{(i)}=H^{(i)} /\sqrt{\rt},\qquad
 R^{(2)}=\sqrt{\rt}L^{(2)},
\\
 L^{(2)}=-\xi^2H^{(2)},\qquad
 S=-\frac{H^{(1)}H^{(2)}}{p\cK},
  \end{gathered}
 \label{a:branch-B}
\end{equation}
where $\xi^2=\frac{3\ii\Omega}{\cK}$.

\subsection{Maxwell and scalar field equations and perturbative accuracy}
\label{a:pdes}

As noted above, our spinor coefficients $P,Q,R$ differ in normalization and chirality convention from the coefficients bearing the same names in the nonrelativistic type-II study~\cite{Keeler2024}. Moreover, for Maxwell fields and zeroth copy scalars truncated at a given perturbative order, we require only that the residuals of the field equations vanish through the retained gradient orders. The coefficients of discarded higher-order terms need not vanish identically.

Eqs.~\KeelerMongaEq{4.14}-\KeelerMongaEq{4.21} of Ref.~\cite{Keeler2024} provide the full component form of Maxwell's equations for general $P,Q,R$ fields. With the background pressure $p$ frozen, the coordinate rescalings $t=p\tau$ and $\rho=\rt/p^2$ bring our background metric~\eqref{a:background} into the form $-\rho\dd t^2+2\dd t\dd\rho+\dd x^2+\dd y^2$, which matches the background metric in Eq.~\KeelerMongaEq{3.9} of Ref.~\cite{Keeler2024} at $\lambda=1$. Thus, rather than re-expanding the exterior derivative and divergence components from scratch, we can directly translate those differential operators into our conventions via Eqs.~\eqref{a:correct-phi} and~\eqref{a:PHL}.

Introducing the transverse complex coordinate $z=x+\ii y$ and the associated derivatives $\partial_z=(\partial_x-\ii\partial_y)/2$ and $\partial_{\bar z}=(\partial_x+\ii\partial_y)/2$, the source-free Maxwell equations on this background are equivalent to the four complex differential equations
\begin{equation}
 \begin{aligned}
 H_r&=-\ii pP_z,&
 &(\rt L)_r+2pL_\tau=-\ii pP_{\bar z},\\
 P_r&=-4\ii pL_z,&
 &P_\tau=-2\ii(H_{\bar z}-\rt L_z),
 \end{aligned}
 \label{a:full-Maxwell}
\end{equation}
which are fully equivalent to $\dd F=0$ and $\bar\nabla_\mu F^{\mu\nu}=0$ for the real field strength $F$. Subscripts \(r,\tau,z,\bar z\) on the scalar functions denote partial derivatives with respect to the indicated coordinates. For branch A, where $L^{(i)}=0$, Eq.~\eqref{a:full-Maxwell} simplifies to
\begin{equation}
 P_r=0,\qquad P_{\bar z}=0,\qquad
 H_r=-\ii pP_z,\qquad P_\tau=-2\ii H_{\bar z}.
 \label{a:A-Maxwell}
\end{equation}
On the other hand, the zeroth copy scalar operator can be written as
\begin{equation}
 \bar\Box S=\frac2p S_{\tau r}
 +\frac1{p^2}\partial_r(\rt S_r)+\Delta_\perp S.
 \label{a:scalar}
\end{equation}
Here $\Delta_\perp\equiv\partial_x^2+\partial_y^2
=4\partial_z\partial_{\bar z}$ denotes the transverse Laplacian.
Whenever the scalar field is independent of the radial coordinate
($S_r=0$), Eq.~\eqref{a:scalar} reduces to $\bar\Box S=\Delta_\perp S$, to which purely time-dependent terms do not contribute.
Whether an additional differential constraint must be imposed on the transverse residual $\Delta_\perp S$ depends on the gradient order assigned to $S$ and on the target perturbative accuracy.

In a local Petrov type-II region where $\Omega,\cK\sim\ord{\partial}$, we choose slowly varying amplitudes $H^{(i)}\sim\ord{\partial}$ for branch A in Eq.~\eqref{a:branch-A}. The algebraic relations~\eqref{a:branch-A} then imply $\xi\sim\ord{1}$, $P^{(i)}\sim\ord{\partial}$, and $S\sim\ord{\partial}$. The radial equations in Eq.~\eqref{a:A-Maxwell} require $P^{(i)}_r,H^{(i)}_r\sim\ord{\partial^2}$, meaning that their leading first-order coefficients are independent of the radial coordinate $r$. Choosing radially independent first-order amplitudes and a radially independent zeroth copy scalar, the Maxwell fields, the zeroth copy scalar, and their boundary derivatives have the following gradient orders
\begin{equation}
 \begin{aligned}
&H^{(i)}\sim\ord{\partial},\qquad
\xi\sim\ord{1},\qquad
P^{(i)}\sim\ord{\partial},\qquad
S\sim\ord{\partial},\\
&\partial_aP^{(i)},\ \partial_aH^{(i)}
\sim\ord{\partial^2},
\qquad a=\tau,x,y,
\end{aligned}
\label{a:slow-orders}
\end{equation}
where $\partial_a\xi\sim\ord{\partial}$ follows from the local nondegeneracy conditions and the gradient counting. Substituting the order assignments~\eqref{a:slow-orders} into Eq.~\eqref{a:A-Maxwell} shows that all residuals of the Maxwell equations begin at second order in gradients. Likewise, Eq.~\eqref{a:scalar} shows that the radially independent scalar field satisfies $\bar\Box S=\Delta_\perp S$, whose residual begins at third order in gradients:
\begin{equation}
\dd F^{(i)}\sim\ord{\partial^2},\qquad
\bar\nabla_\mu F^{(i)\mu\nu}\sim\ord{\partial^2},\qquad
\bar\Box S\sim\ord{\partial^3}.
\label{a:slow-residuals}
\end{equation}
Consequently, under the local nondegeneracy conditions, gradient order assignments for slowly varying amplitudes, and fixed flat Rindler background specified above, branch A provides a consistent Petrov type-II mixed single-copy construction at first order in the gradient expansion.

By contrast, for branch B in Eq.~\eqref{a:branch-B}, we have $P^{(2)}=0$ and $L^{(2)}=-\xi^2H^{(2)}$. Eq.~\eqref{a:full-Maxwell} requires $L^{(2)}_r\sim\ord{\partial^2}$, while the gradient counting gives $L^{(2)}_\tau\sim\ord{\partial^2}$. On the frozen background, however, $\partial_r\rt=p^2$ holds identically, so the second equation in Eq.~\eqref{a:full-Maxwell}, $(\rt L^{(2)})_r+2pL^{(2)}_\tau=-\ii pP^{(2)}_{\bar z}=0$, becomes
\begin{equation}
\underbrace{p^2L^{(2)}}_{\text{possibly first order}}
+\underbrace{\rt L^{(2)}_{r}}_{\text{at least second order}}
+\underbrace{2p  L^{(2)}_\tau}_{\text{at least second order}}
\sim\ord{\partial^2}.
\label{a:branch-B-order}
\end{equation}
Because the last two terms are at least of order $\ord{\partial^2}$, they cannot cancel a nonzero first-order contribution from $p^2L^{(2)}$, forcing $L^{(2)}\sim\ord{\partial^2}$. Branch B therefore cannot support a nondegenerate first-order Petrov type-II single-copy construction under the present assignment of gradient orders and fixed flat Rindler background.

\subsection{Single-copy fields with symmetric magnetic normalization}
\label{a:example}

To make direct contact with the purely magnetic Petrov type-D double copy in 
Ref.~\cite{RelDraft}, we adopt a symmetric magnetic normalization in branch A such that the two single-copy fields share the same $P$ coefficient while carrying opposite $Q$ coefficients. With $c$ denoting a nonzero real constant, the algebraic system~\eqref{a:five-algebraic} yields
\begin{equation}
\begin{aligned}
P^{(1)}&=P^{(2)}=\ii c\cB,\qquad H^{(1)}=-H^{(2)}=H,\\
\qquad
H^2&=\frac{\ii c^2p\cB\cK}{3},\qquad \quad \ S=\frac{\ii c^2\cB}{3},
\end{aligned}
 \label{a:symmetric-copy}
\end{equation}
where the square-root branch for $H$ is chosen consistently with Eq.~\eqref{a:branch-A} so that $\xi H=\ii c\cB$. Setting $Q=H/\sqrt{\rt}$, we have $Q^{(1)}=Q$ and $Q^{(2)}=-Q$. Denoting $F^+=F^{(1)}$ and $F^-=F^{(2)}$, the real field strength expression~\eqref{a:full-F} gives the two single-copy two-forms
\begin{equation}
 F^\pm
=c\cB \dd x\wedge \dd y\mp\sqrt{\rt} \dd\tau\wedge
[\ImPart Q \dd x+\RePart Q \dd y].
 \label{a:symmetric-F}
\end{equation}
The two single-copy fields thus share an identical spatial magnetic component $c\cB \dd x\wedge\dd y$, governed solely by the vorticity-dependent amplitude $\cB$, whereas their radiative components have equal magnitude and opposite sign, with amplitudes determined jointly by the vorticity-dependent amplitude $\cB$ and complex shear $\cK$. A mixed double copy for a general Petrov type-II fluid geometry therefore necessarily involves nonvanishing radiative components, reducing to a purely magnetic Petrov type-D configuration only when the fluid shear vanishes.

Under the local nondegeneracy conditions and assumptions of slow variation of subsection~\ref{a:pdes}, $\cB,\cK\sim\ord{\partial}$ implies $H\sim\ord{\partial}$, so all coefficients in Eq.~\eqref{a:symmetric-copy} obey the assignments of gradient orders~\eqref{a:slow-orders}. The Maxwell and zeroth copy scalar residuals then automatically satisfy the estimates~\eqref{a:slow-residuals}, allowing both the fluid vorticity and shear to vary slowly in space and time at first order. If, within a fixed local domain where $\rt>0$, we keep $\cB\ne0$ while taking the shear-free limit $\cK\to0$, then $H,Q\to0$: the radiative terms with opposite signs in the two single-copy fields vanish simultaneously, and both factors coalesce into the same purely magnetic Petrov type-D single-copy field.

\section{Continuous degeneration of the type-II double copy }
\label{ch:degeneration-recalculated}

This section builds on the relativistic type-II double copy in section~\ref{ch:construction}. It examines whether the family of Petrov type-II mixed single-copy solutions can degenerate along continuous parameter paths to the Petrov type-D and type-N single-copy configurations established in the first-order relativistic study~\cite{RelDraft}. It also examines whether the Maxwell fields and the zeroth-copy scalar continue to satisfy their field equations to the required perturbative accuracy along the entire degeneration path at finite, nonzero values of the degeneration parameter.

\subsection{Degeneration paths and perturbative accuracy requirements}
\label{c3n:setup}

For branch A in section~\ref{ch:construction}, at finite values of the degeneration parameter, the two single-copy spinors and the zeroth copy scalar are determined by Eq.~\eqref{a:branch-A}, the real electromagnetic field strengths are given by Eq.~\eqref{a:full-F}, and the Maxwell and scalar field equations are given by Eqs.~\eqref{a:A-Maxwell} and~\eqref{a:scalar}. To distinguish the perturbative gradient expansion cleanly from the degeneration of algebraic Petrov type, we use the small parameter $\delta$ to track the order of the relativistic gradient expansion introduced above and introduce an independent dimensionless external parameter $s\in[0,1]$ to parametrize the degeneration process. We consider the following two degeneration paths:
\begin{equation}
\begin{aligned}
\text{Type-D path:}\quad&
\cB_s=\cB_0, &\cK_s=s\cK_0,\\
\text{Type-N path:}\quad&
\cB_s=s\cB_0, &\cK_s=\cK_0,
\end{aligned}
\label{c3n:degeneration-paths}
\end{equation}
where $\cB_0$ and $\cK_0$ are nonzero in the chosen local domain and obey the standard first-order conditions on slowly varying fields. From the first-order Weyl scalars in Eq.~\eqref{a:weyl}, taking $s\to0$ along these two paths removes the shear contribution ($\Psi_4\to0$) and the vorticity contribution ($\Psi_2\to0$), respectively, thereby realizing a continuous geometric degeneration from Petrov type II to Petrov type D and Petrov type N.

While the two paths above drive the first-order Weyl curvature continuously from Petrov type II to types D and N, a consistent double copy degeneration requires more than algebraic convergence. The single-copy fields must recover the known type-D/N configurations in the limit $s\to0$ and simultaneously satisfy the Maxwell and scalar field equations to the prescribed accuracy at every finite $s>0$. In particular, the higher-order residuals permitted in the field equations must remain uniformly controlled throughout the deformation and must not be singularly amplified as $s\to0$, which would otherwise invalidate the underlying perturbative approximation.

This uniformity requirement restricts the freedom to rescale the single-copy field amplitudes. Although a simultaneous rescaling $f^{(i)}\to a_i f^{(i)}$ and $S\to a_1a_2S$ leaves the algebraic double copy relation and the reconstructed Weyl curvature invariant, spacetime-dependent factors $a_i$ alter the spacetime gradients of the electromagnetic fields and introduce additional derivative terms into the Maxwell equations:
\begin{equation}
\bar\nabla^{AA'}\left(a_i f^{(i)}_{AB}\right)
=a_i\bar\nabla^{AA'}f^{(i)}_{AB}+
f^{(i)}_{AB}\bar\nabla^{AA'}a_i.
\label{c3n:rescaling-Maxwell}
\end{equation}
Whether the rescaled fields continue to meet the prescribed perturbative accuracy must therefore be checked explicitly for each normalization choice.

\subsection{Shear-free limit and the type-D double copy}
\label{c3n:D-section}

For the shear-free degeneration, we adopt the symmetric magnetic normalization of subsection~\ref{a:example}, given in Eq.~\eqref{a:symmetric-copy}, with $c=1$. The spinor coefficients $P^{(1)}$ and $P^{(2)}$ as well as the zeroth copy scalar $S_s$ are then independent of the degeneration parameter $s$, whereas the shear-dependent radiative amplitude scales as
\begin{equation}
H_s=\sqrt{s} H_0,
\qquad
H_0^2=\frac{\ii p\cB_0\cK_0}{3}.
\label{c3n:D-amplitude}
\end{equation}
Accordingly, $H_0\sim\ord{\delta}$ and $\partial_aH_0\sim\ord{\delta^2}$. Since the external parameter $s$ is independent of the spacetime coordinates, Eq.~\eqref{c3n:D-amplitude} implies $\partial_aH_s\sim\ord{\sqrt{s}\delta^2}$, while the remaining coefficients retain their original gradient orders from section~\ref{ch:construction}. Consequently, the $\ord{\delta^2}$ Maxwell residuals and the $\ord{\delta^3}$ zeroth copy scalar residual in Eq.~\eqref{a:slow-residuals} hold uniformly in $s$ along the entire type-D path ($0\le s\le1$).

In the degeneration limit $s\to0$, the shear-induced radiative terms with opposite signs in the two single-copy fields vanish simultaneously as $\sqrt{s}$, and the two inequivalent type-II factors coalesce into a single Petrov type-D factor:
\begin{equation}
\begin{aligned}
f^{(1)}_{AB}, f^{(2)}_{AB}
&\longrightarrow
f_{D AB}
=\ii\cB_0 o_{(A}\iota_{B)},\\
S_s&\longrightarrow S_D=\frac{\ii\cB_0}{3},\\
F^{(1)}, F^{(2)}
&\longrightarrow F_D=\cB_0 \dd x\wedge \dd y.
\end{aligned}
\label{c3n:D-endpoint}
\end{equation}
This limit reproduces the purely magnetic Petrov type-D single-copy configuration of Eqs.~\WangEtAlEq{45}-\WangEtAlEq{51} in Ref.~\cite{RelDraft} for $c_B=1$ and $\theta=0$. Throughout this process, the vorticity-dependent amplitude $\cB_0$ need not be restricted to a spacetime constant. It is only required to obey the first-order conditions on slowly varying fields, with second- and higher-order Maxwell residuals permitted in accordance with the discussion of first-order slowly varying amplitudes following Eq.~\WangEtAlEq{58} in Ref.~\cite{RelDraft}.

\subsection{Irrotational limit and the type-N double copy}
\label{c3n:N-section}

For the type-N degeneration, retaining the symmetric magnetic normalization of Eq.~\eqref{a:symmetric-copy} would cause both single-copy fields and the zeroth copy scalar to vanish identically as the vorticity approaches zero, so a naive substitution of $\cB=0$ with that normalization cannot yield a nontrivial Petrov type-N single copy. To preserve a nonvanishing radiative amplitude in the irrotational limit, we return to the general parametrization of branch A in Eq.~\eqref{a:branch-A} and adopt the $\beta$-dependent amplitude normalization introduced in Eqs.~\WangEtAlEq{70}-\WangEtAlEq{71} of the first-order relativistic type-D/N study~\cite{RelDraft}:
\begin{equation}
H^{(1)}=H^{(2)}=\mathcal A_\beta,
\qquad
\mathcal A_\beta
=c_\beta e^{\ii\theta}\cK_0^\beta,
\qquad \beta=0,1,
\label{c3n:N-amplitude}
\end{equation}
where $c_\beta$ is a nonzero real constant and $\theta$ is a constant real phase. The remaining spinor coefficients and the zeroth copy scalar are then fixed by Eq.~\eqref{a:branch-A}. Fixing a single-valued branch of the square root, the type-N path yields
\begin{equation}
\xi_s=\sqrt{s} \xi_*,
\qquad
\xi_*^2=\frac{3\ii\cB_0}{p\cK_0}.
\label{c3n:N-root-scaling}
\end{equation}
As $s\to0$, the coefficients of the mixed terms $P^{(1)}$ and $P^{(2)}$ therefore vanish as $\sqrt{s}$, whereas the two radiative $Q$ components retain a common nonzero limit. From Eq.~\eqref{a:branch-A}, the limiting configuration as $s\to0$ is
\begin{equation}
\begin{aligned}
f^{(1)}_{AB}, f^{(2)}_{AB}
&\longrightarrow
f_{N_\beta AB}
=\frac{\mathcal A_\beta}{\sqrt{\rt}} o_Ao_B,\\
S_s&\longrightarrow
S_{N_\beta}
=-\frac{c_\beta^2e^{2\ii\theta}}p 
\cK_0^{ 2\beta-1}.
\end{aligned}
\label{c3n:N-endpoint}
\end{equation}
Eq.~\eqref{c3n:N-endpoint} matches the Petrov type-N normalization in Eqs.~\WangEtAlEq{70}-\WangEtAlEq{71} of Ref.~\cite{RelDraft}, from which the corresponding real field strength and algebraic curvature reconstruction follow directly. At the spinor level, as the vorticity tends to zero, the mixed components proportional to $o_{(A}\iota_{B)}$, which have opposite signs, in the two type-II spinor factors vanish simultaneously, while the remaining radiative components $o_Ao_B$ converge to a common Petrov type-N single-copy factor describing a null electromagnetic field. Here the parameter $\beta$ governs how the fluid shear is partitioned between the gauge and scalar sectors: for $\beta=1$ (the $N_1$ branch), the single-copy electromagnetic field amplitude is proportional to the complex shear $\cK_0$, whereas for $\beta=0$ (the $N_0$ branch), the single-copy radiative amplitude is held fixed and the entire dependence on the fluid shear is carried by the zeroth copy scalar field $S_{N_0}\propto\cK_0^{-1}$, in full agreement with Ref.~\cite{RelDraft}.

\subsection{\texorpdfstring{Additional conditions for $N_1$ and $N_0$ degenerations}{Residual estimates and additional conditions for N1 and N0 degenerations}}
\label{c3n:N-uniform}

With the amplitude choice~\eqref{c3n:N-amplitude}, we make the gradient expansion parameter $\delta$ explicit in the slowly varying kinematic data by writing
\begin{equation}
\cB_s=s\delta b,\qquad
\cK_0=\delta\kappa,
\qquad 0<s\le1,
\label{c3n:N-slow-data}
\end{equation}
where the rescaled amplitudes $b$ and $\kappa$ are independent of the degeneration parameter $s$.

For $\beta=1$, combining Eqs.~\eqref{c3n:N-amplitude}, \eqref{c3n:N-root-scaling}, and~\eqref{a:branch-A} gives the gradient orders of the Maxwell fields and zeroth copy scalar
\begin{equation}
\mathcal A_1\sim\ord{\delta},\qquad
P^{(1)},P^{(2)}\sim\ord{\sqrt{s}\delta},\qquad
S_s\sim\ord{\delta}.
\label{c3n:N1-orders}
\end{equation}
Because the fluid data vary slowly along the boundary directions, each boundary derivative raises the gradient order by one. Furthermore, since the degeneration parameter $s$ is independent of the spacetime coordinates, the suppression factor $\sqrt{s}$ in $P^{(1)}$ and $P^{(2)}$ is preserved under differentiation. Consequently, as the vorticity diminishes, the Maxwell fields, the zeroth copy scalar, and their derivatives maintain the gradient orders specified in Eq.~\eqref{a:slow-orders}, and the residuals of the Maxwell equations and the zeroth copy scalar equation remain $\ord{\delta^2}$ and $\ord{\delta^3}$, respectively, without developing any singularity as $s\to0$. The original perturbative accuracy is therefore preserved uniformly along the entire $N_1$ degeneration path. In particular, when the kinematic data in Eq.~\eqref{c3n:N-slow-data} arise from a family of fluid solutions satisfying the first-order fluid constraints~\eqref{a:metric-euler} along the path, this deformation describes a physical fluid family whose vorticity gradually vanishes at finite shear, and the Petrov type-II single copy deforms continuously into the $N_1$ configuration while maintaining the required perturbative accuracy throughout.

For $\beta=0$, the radiative amplitude $\mathcal A_0=c_0e^{\ii\theta}$ is a fixed nonzero constant, shifting the gradient orders of the Maxwell fields and zeroth copy scalar to
\begin{equation}
\mathcal A_0\sim\ord{1},\qquad
P^{(1)},P^{(2)}\sim\ord{\sqrt{s}},\qquad
S_s\sim\ord{\delta^{-1}}.
\label{c3n:N0-orders}
\end{equation}
This order assignment lies outside the direct scope of Eq.~\eqref{a:slow-residuals}. Indeed, generic first-order conditions on slowly varying fields guarantee only
\begin{equation}
\partial_aP^{(i)}
\sim\ord{\sqrt{s}\delta},
\qquad a=\tau,x,y,
\label{c3n:N0-generic-derivatives}
\end{equation}
so at finite $s>0$ the Maxwell equations generically retain a first-order residual of $\ord{\sqrt{s}\delta}$. To ensure that the Maxwell residuals remain uniformly $\ord{\delta^2}$ across the entire interval $0<s\le1$, all boundary derivatives of $P^{(i)}$ must be of second order in the gradient expansion, which requires
\begin{equation}
\partial_a\xi_*\sim\ord{\delta^2}
\quad\Longleftrightarrow\quad
\partial_a\!\left(\frac b\kappa\right)\sim\ord{\delta^2},
\qquad a=\tau,x,y.
\label{c3n:N0-ratio-condition}
\end{equation}
Whenever Eq.~\eqref{c3n:N0-ratio-condition} is not satisfied, this minimal extension does not satisfy Maxwell's equations through first order at finite $s$. Higher-order field corrections must then be included and the degeneration limit re-evaluated.

Similarly, for the zeroth copy scalar field $S_s\sim\ord{\delta^{-1}}$, radial independence implies that applying the scalar wave operator~\eqref{a:scalar} gives
\begin{equation}
\bar\Box S_s
=
-\frac{c_0^2e^{2\ii\theta}}p 
\Delta_\perp(\cK_0^{-1})
\sim\ord{\delta},
\label{c3n:N0scalar-residual}
\end{equation}
which leaves an $\ord{\delta}$ residual rather than the $\ord{\delta^3}$ residual estimate of Eq.~\eqref{a:slow-residuals}. Eliminating this first-order residual in the zeroth copy scalar equation requires imposing the additional condition on the transverse Laplacian
\begin{equation}
\Delta_\perp(\cK_0^{-1})
\sim\ord{\delta^2},
\label{c3n:N0scalar-condition}
\end{equation}
namely that the leading term of the transverse Laplacian acting on the inverse complex shear $\cK_0^{-1}$ vanish. If one further demands the $\ord{\delta^3}$ scalar residual estimate of Eq.~\eqref{a:slow-residuals}, the stronger condition $\Delta_\perp(\cK_0^{-1})\sim\ord{\delta^3}$ must be imposed.

\section{Nonrelativistic limit of the type-II double copy}
\label{b:chapter}

We now turn to the nonrelativistic limit of the type-II double copy. We first analyze the transformations of the background metric, null coframe, and spinor coefficients, and then compare our results with 
the nonrelativistic type-II double copy study~\cite{Keeler2024}, covering the leading-order Weyl curvature, the electromagnetic fields of the two single-copy factors, the zeroth copy scalar field, and the functional dependence of the amplitudes.

\subsection{Nonrelativistic hydrodynamic scaling}
\label{b:scaling}

The incompressible nonrelativistic limit of relativistic hydrodynamics and its gravitational realization were established in Ref.~\cite{BhattacharyyaMinwallaWadia2009}. Following the nonrelativistic hydrodynamic scaling expansion in Eqs.~\WangEtAlEq{59}-\WangEtAlEq{61} of section~\href{https://arxiv.org/pdf/2609.29922v1\#nameddest=section*.17}{V.C} of Ref.~\cite{RelDraft}, we set the cutoff radius and the equilibrium background pressure to $r_c=p_0=1$ and introduce slow nonrelativistic coordinates together with the scaling parameter $\lambda$,
\begin{equation}
 T=\epsilon^2\tau,\qquad X=\epsilon x,\qquad Y=\epsilon y,
 \qquad \lambda=\epsilon^2,
 \label{b:slow-coordinates}
\end{equation}
expanding the boundary fluid velocity and pressure in the small parameter $\epsilon$ as
\begin{equation}
 u_i=\epsilon v_i(T,X,Y)+\ord{\epsilon^3},\qquad
 p=1+\epsilon^2\Pi(T,X,Y)+\ord{\epsilon^4}.
 \label{b:slow-fluid}
\end{equation}
Under this scaling, a spatial derivative acting on a slowly varying function introduces one power of $\epsilon$, whereas a time derivative introduces two powers, $\epsilon^2$. Here $p$ denotes the hydrodynamic pressure on the gravitational side. The Maxwell and scalar field equations on the gauge theory side remain formulated on the frozen background of Eq.~\eqref{a:background}, specialized here to $p_0=1$.

Imposing the leading-order incompressibility condition $\partial_Xv_X+\partial_Yv_Y=0$, we define the nonrelativistic vorticity $\omega$ and complex shear $\kappa$ by
\begin{equation}
 \omega=\frac12(\partial_Xv_Y-\partial_Yv_X),\qquad
 \kappa=\partial_Xv_X-\frac{\ii}{2}
 (\partial_Xv_Y+\partial_Yv_X).
 \label{b:omega-kappa}
\end{equation}
Substituting Eqs.~\eqref{b:slow-coordinates} and~\eqref{b:slow-fluid} into the definitions of the relativistic fluid kinematic quantities in section~\ref{ch:construction} yields
\begin{equation}
 \cB=\epsilon^2\omega+\ord{\epsilon^4},\qquad
 \cK=\epsilon^2\kappa+\ord{\epsilon^4},\qquad \rt=r+\ord{\epsilon^2}.
 \label{b:data-limit}
\end{equation}
The vorticity and shear contributions to the first-order Weyl curvature in section~\ref{ch:construction} thus both enter at $\ord{\epsilon^2}$ in the nonrelativistic limit, with their leading-order nonrelativistic data determined by $\omega$ and $\kappa$, respectively. Note that the order in the relativistic gradient expansion cannot simply be identified with the power of $\epsilon$ in the nonrelativistic scaling expansion. For instance, a second-order relativistic gradient term such as $\partial_iK_{jk}$ is of order $\ord{\epsilon^3}$ under this scaling and therefore cannot, in general, be absorbed into an $\ord{\epsilon^4}$ remainder.

\subsection{Rescaling and frame transformations}
\label{b:frame}

Here the superscript $(i)$ ($i=1,2$) distinguishes the two single-copy factors, while the superscript $0$ labels the leading-order coefficient in the $\epsilon$ expansion. We expand the relativistic quantities of section~\ref{ch:construction} in powers of $\epsilon$ as
\begin{equation}
 \begin{aligned}
 P^{(i)}&=\epsilon^2P_R^{(i),0}+\ord{\epsilon^3},\quad H^{(i)}=\epsilon^2H_R^{(i),0}+\ord{\epsilon^3}, \quad S=\epsilon^2S_R^0+\ord{\epsilon^3},
 \end{aligned}
 \label{b:leading-amplitudes}
\end{equation}
where $H^{(i)}$ retains its definition in Eq.~\eqref{a:PHL}, the subscript $R$ indicates quantities evaluated in the relativistic frame prior to the frame transformation, and $P^{(i),0},Q^{(i),0},R^{(i),0},S^0$ denote the corresponding leading-order coefficients after transformation to the target nonrelativistic frame.

The fixed flat Rindler background of section~\ref{ch:construction}, given in Eq.~\eqref{a:background}, takes the general form
\begin{equation}
\dd\bar s^2=-\rt \dd\tau^2+2p \dd\tau \dd r+\dd x^2+\dd y^2.
\label{b:original-background}
\end{equation}
Setting the frozen background pressure to $p_0=1$ gives $\rt=r$, reducing the background metric to
\begin{equation}
\dd\bar s^2=-r \dd\tau^2+2 \dd\tau \dd r+\dd x^2+\dd y^2.
\label{b:unit-background}
\end{equation}
For branch A ($R^{(i)}=0$), Eqs.~\eqref{a:full-F} and~\eqref{a:PHL} give the real field strength two-form
\begin{equation}
F^{(i)}=\RePart P^{(i)} \dd\tau\wedge \dd r
+\ImPart P^{(i)} \dd x\wedge \dd y
-\dd\tau\wedge
\bigl(\ImPart H^{(i)} \dd x+\RePart H^{(i)} \dd y\bigr).
\label{b:branch-A-F}
\end{equation}
In terms of the slow nonrelativistic coordinates~\eqref{b:slow-coordinates}, the coordinate differentials are
\begin{equation}
\dd\tau=\frac{\dd T}{\epsilon^2},\qquad
\dd x=\frac{\dd X}{\epsilon},\qquad
\dd y=\frac{\dd Y}{\epsilon}.
\label{b:coordinate-differentials}
\end{equation}
Let $\Phi_\epsilon^*$ denote the pullback rewriting tensors from the original coordinates in the rescaled coordinates $(T,r,X,Y)$. The pulled-back background metric is then
\begin{equation}
\Phi_\epsilon^*\bar g
=-\frac r{\epsilon^4}\dd T^2
+\frac2{\epsilon^2}\dd T \dd r
+\frac1{\epsilon^2}(\dd X^2+\dd Y^2).
\label{b:pullback-background}
\end{equation}
To adopt the same overall normalization as the background metric in Eq.~\KeelerMongaEq{3.9} of the nonrelativistic type-II study~\cite{Keeler2024}, we define the conformally rescaled background metric and pulled-back field strength by
\begin{equation}
 \hat g=\epsilon^2\Phi_\epsilon^*\bar g,
 \qquad \hat F=\Phi_\epsilon^*F.
 \label{b:metric-field-map}
\end{equation}
Under the constant conformal rescaling of the metric, we assign zero conformal weight to the field-strength two-form $F$, as permitted by the conformal invariance of the four-dimensional source-free Maxwell equations. The rescaled background line element is
\begin{equation}
 \begin{aligned}
 \dd\hat s^2&=-\frac r{\epsilon^2}\dd T^2+2 \dd T \dd r+\dd X^2+\dd Y^2\\
 &=-\frac r\lambda \dd T^2+2 \dd T \dd r+\dd X^2+\dd Y^2,
 \end{aligned}
 \label{b:hat-background}
\end{equation}
which coincides with the near-horizon background metric in Eq.~\KeelerMongaEq{3.9} of Ref.~\cite{Keeler2024}.

Setting $p_0=1$ in Eq.~\eqref{a:tetrad}, the original relativistic null coframe reads
\begin{equation}
\ell=-\sqrt{\frac r2} \dd\tau,\quad
n=-\sqrt{\frac r2} \dd\tau+\sqrt{\frac2r} \dd r,\quad
m=\frac{-\ii \dd x+\dd y}{\sqrt2}, \quad
\bar m=\frac{\mathrm i\mathrm dx+\mathrm dy}{\sqrt2}.
\label{b:unit-R-frame}
\end{equation}
Since Eq.~\eqref{b:metric-field-map} multiplies the background metric by $\epsilon^2$, the corresponding null coframe one-forms must be rescaled by $\epsilon$ after the coordinate pullback. Denoting the rescaled relativistic null coframe by $(L,N,M)$, we obtain
\begin{equation}
 \begin{aligned}
 L&=-\frac{\sqrt r}{\sqrt2\epsilon}\dd T,&
 N&=-\frac{\sqrt r}{\sqrt2\epsilon}\dd T
       +\epsilon\sqrt{\frac2r}\dd r,\\
 M&=\frac{-\ii\dd X+\dd Y}{\sqrt2}.
 \end{aligned}
 \label{b:rescaled-R-frame}
\end{equation}
We refer to the null coframe corresponding to Eqs.~\KeelerMongaEq{A.9} and~\KeelerMongaEq{A.13} in the appendix of Ref.~\cite{Keeler2024} as the $K$ frame:
\begin{equation}
 \begin{aligned}
 \ell_K&=\frac{\dd T}{\sqrt2\epsilon}=-r^{-1/2}L,\\
 n_K&=\frac r{\sqrt2\epsilon}\dd T-\sqrt2\epsilon\dd r
       =-r^{1/2}N,\\
 m_K&=\frac{\ii\dd X+\dd Y}{\sqrt2}=\bar M.
 \end{aligned}
 \label{b:K-frame}
\end{equation}
The $R\to K$ frame transformation in Eq.~\eqref{b:K-frame} is the composition of three operations: a radial Lorentz boost, a simultaneous orientation reversal of the two real null directions, and the interchange $M\leftrightarrow\bar M$ by complex conjugation of the transverse complex null directions. Both the Maxwell and Weyl spinor components transform according to this composite frame map.

In what follows, we work consistently in the $K$ frame and maintain the correspondence between the NP scalars and spinor components through Eqs.~\eqref{a:correct-phi} and~\eqref{a:weyl-spinor}. Let $(o_A,\iota_A)$ denote the spin dyad adapted to the $K$ frame, normalized by $o_A\iota^A=1$. The leading-order Maxwell spinors expand in the $K$ frame as
\begin{equation}
 f_{AB}^{(i),0}=P^{(i),0}o_{(A}\iota_{B)}
 +Q^{(i),0}o_Ao_B+R^{(i),0}\iota_A\iota_B,
\label{b:leading-spinor-expansion}
\end{equation}
with $\phi_0=R$, $\phi_1=-P/2$, and $\phi_2=Q$ unchanged in form. Under the constant conformal rescaling $\hat g=\epsilon^2\Phi_\epsilon^*\bar g$, the frame components of the Maxwell field strength acquire a scaling weight of $\epsilon^{-2}$, and those of the target Weyl curvature on the gravitational side acquire the same weight $\epsilon^{-2}$. Normalizing the zeroth copy scalar as $\epsilon^{-2}S$ therefore preserves the scaling invariance of the algebraic Weyl double copy relation.

To derive the explicit transformation laws for the expansion coefficients, we substitute Eqs.~\eqref{b:leading-amplitudes} and~\eqref{b:coordinate-differentials} into Eq.~\eqref{b:branch-A-F}. Expressed in terms of the original coefficients in the $R$ frame, the pulled-back real field strength is
\begin{equation}
 \begin{aligned}
 \hat F^{(i)}
={}&\RePart P_R^{(i),0} \dd T\wedge \dd r
+\ImPart P_R^{(i),0} \dd X\wedge \dd Y\\
&-\frac1\epsilon \dd T\wedge
\bigl(\ImPart H_R^{(i),0} \dd X+\RePart H_R^{(i),0} \dd Y\bigr).
 \end{aligned}
 \label{b:pullback-F}
\end{equation}
In the $K$ frame, under our NP conventions, the same physical real field strength two-form takes the form
\begin{equation}
\begin{aligned}
\hat F^{(i)}
={}&\RePart P^{(i),0} \dd T\wedge \dd r
-\ImPart P^{(i),0} \dd X\wedge \dd Y\\
&+\frac1\epsilon \dd T\wedge
\bigl(-\ImPart Q^{(i),0} \dd X+\RePart Q^{(i),0} \dd Y\bigr).
\end{aligned}
\label{b:K-F}
\end{equation}
Matching the real coefficients of the linearly independent coordinate two-forms $\dd T\wedge\dd r$, $\dd X\wedge\dd Y$, $\dd T\wedge\dd X$, and $\dd T\wedge\dd Y$ term by term yields
\begin{align}
\RePart P^{(i),0}&=\RePart P_R^{(i),0},\qquad
\ImPart P^{(i),0}=-\ImPart P_R^{(i),0},
\label{b:P-real-map}
\\
\RePart Q^{(i),0}&=-\RePart H_R^{(i),0},\quad
\ImPart Q^{(i),0}=\ImPart H_R^{(i),0}.
\label{b:Q-real-map}
\end{align}
Combining the transformation of the Weyl scalars under this frame map with the double copy relation of section~\ref{ch:construction}, these real and imaginary relations can be written compactly as the transformation laws involving complex conjugation
\begin{equation}
 P^{(i),0}=\overline{P_R^{(i),0}},\qquad Q^{(i),0}=-\overline{H_R^{(i),0}},
 \qquad R^{(i),0}=0,\qquad S^0=\overline{S_R^0}.
 \label{b:coeff-map}
\end{equation}
Similarly, denoting by $\Psi_{2R},\Psi_{4R}$ the leading-order Weyl scalars after multiplication by $\epsilon^{-2}$ induced by the constant metric rescaling in Eq.~\eqref{b:metric-field-map} but prior to transforming to the $K$ frame, the frame relation~\eqref{b:K-frame} gives
\begin{equation}
 \Psi_{2K}=\overline{\Psi_{2R}},\qquad
 \Psi_{4K}=r\overline{\Psi_{4R}}.
 \label{b:weyl-map}
\end{equation}

Finally, let $P_{\mathrm{pap}}^{(i),0},Q_{\mathrm{pap}}^{(i),0}$ denote the leading coefficients of the two Maxwell spinors in Eq.~\KeelerMongaEq{4.32} of Ref.~\cite{Keeler2024}, with the order in $\lambda$ specified by Eq.~\KeelerMongaEq{4.40} therein. Comparing the expressions for the same real field strength two-form in the two spinor conventions yields
\begin{equation}
 P^{(i),0}=-\frac12\overline{P_{\rm pap}^{(i),0}},\qquad Q^{(i),0}=\frac12\overline{Q_{\rm pap}^{(i),0}}.
 \label{b:paper-dictionary}
\end{equation}
These relations reflect the two differences in convention between Ref.~\cite{Keeler2024} and the present work: an overall factor of $1/2$ in the normalization of the field strength and complex conjugation due to the opposite spinor chirality convention associated with the transverse orientation. Transforming the Weyl curvature spinor under the same normalization and chirality convention, the zeroth copy scalar on the present branch transforms as $(S^0)_{\mathrm{pap}\to K}=\overline{S_{\mathrm{pap}}^0}/4$.

\subsection{Comparison of the leading-order double copy structures}
\label{b:general-copy}

Choosing leading-order radiative amplitudes that depend only on the slow time coordinate $T$,
\begin{equation}
Q^{(i),0}=Q^{(i),0}(T),\qquad i=1,2,
\label{b:time-amplitudes}
\end{equation}
substituting the nonrelativistic scaling expansion~\eqref{b:data-limit} into the curvature expressions~\eqref{a:weyl} of section~\ref{ch:construction}, and applying the frame transformation laws~\eqref{b:weyl-map}, we obtain the leading-order nonrelativistic Weyl scalars in the $K$ frame:
\begin{equation}
\Psi_{2K}=-\frac{\ii\omega}{2},
\qquad
\Psi_{4K}=-\overline{\kappa}.
\label{b:leading-curvature}
\end{equation}
Introducing a real stream function $\chi(T,X,Y)$ for the two-dimensional incompressible flow~\cite{Keeler2024},
\begin{equation}
v_X=\partial_Y\chi,\qquad
v_Y=-\partial_X\chi,
\label{b:stream-function}
\end{equation}
and writing the slow complex coordinate as $Z=X+\ii Y$ with derivatives $\partial_Z=(\partial_X-\ii\partial_Y)/2$ and $\partial_{\bar Z}=(\partial_X+\ii\partial_Y)/2$, Eq.~\eqref{b:omega-kappa} gives
\begin{equation}
\begin{aligned}
\Psi_{2K}
&=-\frac{\ii\omega}{2}
=\ii \partial_Z\partial_{\bar Z}\chi,\\
\Psi_{4K}
&=-\overline{\kappa}
=2\ii \partial_{\bar Z}^{ 2}\chi.
\end{aligned}
\label{b:stream-curvature}
\end{equation}
This agrees with the leading-order geometric curvature in Eq.~\KeelerMongaEq{2.6} of Ref.~\cite{Keeler2024}. We next extract the single-copy spinor coefficients. Combining the algebraic relations for branch A in Eq.~\eqref{a:branch-A}, the nonrelativistic expansions~\eqref{b:data-limit} and~\eqref{b:leading-amplitudes}, and the frame transformation~\eqref{b:coeff-map}, we find
\begin{equation}
\begin{aligned}
(\xi^0)^2&=\frac{3\ii\omega}{\kappa},\\
P^{(1),0}&=-\overline{\xi^0} Q^{(1),0},\\
P^{(2),0}&=\overline{\xi^0} Q^{(2),0}.
\end{aligned}
\label{b:leading-P}
\end{equation}
The coefficients of the $P$ components of the two single-copy fields are governed by the same vorticity-to-shear ratio and carry opposite signs. On the same branch, the coefficient transformation~\eqref{b:coeff-map} gives the leading-order zeroth copy scalar field
\begin{equation}
S^0
=-\frac{Q^{(1),0}Q^{(2),0}}{\overline{\kappa}}
=\frac{Q^{(1),0}Q^{(2),0}}{\Psi_{4K}}.
\label{b:leading-S}
\end{equation}
Combining Eqs.~\eqref{b:leading-curvature} and~\eqref{b:leading-P} gives $(\overline{\xi^0})^2=-6\Psi_{2K}/\Psi_{4K}$. Selecting locally a single-valued square-root branch consistent with Eq.~\eqref{b:leading-P}, namely $\overline{\xi^0}=-\ii\sqrt{6\Psi_{2K}/\Psi_{4K}}$, we can write the two leading-order Maxwell spinors as
\begin{equation}
\begin{aligned}
f^{(1),0}_{AB}
&=Q^{(1),0}
\left[
\ii\sqrt{\frac{6\Psi_{2K}}{\Psi_{4K}}} 
o_{(A}\iota_{B)}
+o_Ao_B
\right],\\
f^{(2),0}_{AB}
&=Q^{(2),0}
\left[
-\ii\sqrt{\frac{6\Psi_{2K}}{\Psi_{4K}}} 
o_{(A}\iota_{B)}
+o_Ao_B
\right].
\end{aligned}
\label{b:corrected-leading}
\end{equation}
Eqs.~\eqref{b:corrected-leading} and~\eqref{b:leading-S} satisfy the leading-order algebraic double copy equations~\eqref{a:five-algebraic} of section~\ref{ch:construction} identically. Through the frame and spinor transformations derived above, relativistic branch A thus reproduces, in the nonrelativistic limit, the leading-order nonrelativistic Weyl spinor whose nonzero components are given in Eq.~\eqref{b:stream-curvature}.

We now compare these single-copy coefficients term by term with those reported in the nonrelativistic type-II study~\cite{Keeler2024}. Following the notation of subsection~\ref{b:frame}, we label the nonrelativistic field strength expansion coefficients of Ref.~\cite{Keeler2024} with the subscript $\mathrm{pap}$ and denote the Weyl scalars in Eqs.~\KeelerMongaEq{2.3} and~\KeelerMongaEq{2.6} of that work by $\Psi_{j,\mathrm{NR}}$. Eq.~\eqref{b:stream-curvature} shows that the geometric curvature scalars on the two sides coincide in the $K$ frame:
\begin{equation}
\Psi_{2,\mathrm{NR}}=\Psi_{2K},\qquad
\Psi_{4,\mathrm{NR}}=\Psi_{4K}.
\label{b:paper-curvature-identification}
\end{equation}
With the square-root branch convention fixed above, Eqs.~\KeelerMongaEq{4.55} and~\KeelerMongaEq{4.58} of Ref.~\cite{Keeler2024} take the form
\begin{equation}
\begin{aligned}
P_{\mathrm{pap}}^{(1),0}
&=\ii Q_{\mathrm{pap}}^{(1),0}
\sqrt{\frac{\Psi_{2,\mathrm{NR}}}{\Psi_{4,\mathrm{NR}}}},\\
P_{\mathrm{pap}}^{(2),0}
&=-\ii Q_{\mathrm{pap}}^{(2),0}
\sqrt{\frac{\Psi_{2,\mathrm{NR}}}{\Psi_{4,\mathrm{NR}}}},\\
S_{\mathrm{pap}}^0
&=\frac{
Q_{\mathrm{pap}}^{(1),0}Q_{\mathrm{pap}}^{(2),0}
}{\Psi_{4,\mathrm{NR}}}.
\end{aligned}
\label{b:paper-leading-factors}
\end{equation}
Comparing Eqs.~\eqref{b:corrected-leading} and~\eqref{b:paper-leading-factors} confirms that the two constructions share the same algebraic structure for the leading-order factorization of the Weyl spinor.

Applying the coefficient transformation~\eqref{b:paper-dictionary} to convert the coefficients of Ref.~\cite{Keeler2024} into our conventions for the real field strength and the spinor coefficients in the $K$ frame, we find that the ratios between the products of coefficients of the mixed and radiative terms in Ref.~\cite{Keeler2024} and in our construction are, respectively,
\begin{equation}
\begin{aligned}
\left(\frac{P^{(1),0}P^{(2),0}}
     {Q^{(1),0}Q^{(2),0}}\right)_{\mathrm{pap}\to K}
&={
\frac{\overline{\Psi_{2,\mathrm{NR}}}}
     {\overline{\Psi_{4,\mathrm{NR}}}}
},\\[4pt]
\frac{P^{(1),0}P^{(2),0}}
     {Q^{(1),0}Q^{(2),0}}
&=
6\frac{\Psi_{2K}}{\Psi_{4K}}.
\end{aligned}
\label{b:factor-ratio-comparison}
\end{equation}
When the same radiative amplitudes $Q^{(i),0}$ in the $K$ frame are chosen on both sides, the zeroth copy scalar of Ref.~\cite{Keeler2024}, converted under the same normalization and chirality convention, becomes
\begin{equation}
(S^0)_{\mathrm{pap}\to K}
=
\frac{Q^{(1),0}Q^{(2),0}}
{\overline{\Psi_{4,\mathrm{NR}}}}.
\label{b:paper-leading-S}
\end{equation}
In summary, the leading-order Weyl curvature obtained from our nonrelativistic limit agrees with the geometric result of Ref.~\cite{Keeler2024}. Both constructions share the same leading-order structure for the factorization of the Weyl spinor into two Maxwell spinors, while the differences between the coefficients reported in Ref.~\cite{Keeler2024} and those derived here reduce to a relative factor of $\sqrt{6}$ in the mixed term , together with the complex conjugation associated with the transverse chirality convention.

\section{Conclusions and discussion}
\label{ch:summary}

Using the first-order geometry of relativistic Rindler fluids, we have constructed a Petrov type-II double copy by factorizing the Weyl spinor into two Maxwell spinors and verified the Maxwell and scalar field equations order by order. Under the local nondegeneracy condition and the adopted gradient counting, branch A reproduces the first-order Weyl spinor through the mixed double copy and satisfies the Maxwell and zeroth-copy scalar equations up to residuals of \(\mathcal{O}(\partial^2)\) and \(\mathcal{O}(\partial^3)\), respectively. Branch B, by contrast, is obstructed by radial and temporal differential constraints and cannot support a nondegenerate first-order construction under the same background and assumptions on gradient counting. Adopting a symmetric magnetic normalization further demonstrates that the two single-copy fields share a common magnetic component while carrying equal and opposite radiative components.

In our degeneration analysis, separating the external deformation parameter $s$ from the gradient expansion parameter $\delta$ allowed us to examine the shear-free and irrotational limits on an equal footing. Along the shear-free path, the radiative terms with opposite signs in the two single-copy fields decay as $\sqrt{s}$, coalescing into the purely magnetic type-D double copy. Along the irrotational path, we normalize the two Maxwell spinors so that their common radiative amplitude remains finite and nonzero as \(s\to0\). Their mixed terms then vanish as \(\sqrt{s}\), and both spinors converge to the same type-N single-copy factor. Both limiting configurations reproduce the established relativistic type-D and type-N double copies~\cite{RelDraft}. Under uniform conditions on slowly varying fields, the Maxwell and scalar residuals along the type-D and $N_1$ ($\beta=1$) paths remain $\ord{\delta^2}$ and $\ord{\delta^3}$, respectively, without singular amplification as $s\to0$, so both degenerations preserve the perturbative accuracy of the Maxwell and scalar field equations alongside the algebraic limit. For the fixed-amplitude \(N_0\) branch (\(\beta=0\)), the scalings \(\mathcal A_0=O(1)\), \(P^{(i)}=O(\sqrt{s})\), and \(S_s=O(\delta^{-1})\) require separate residual estimates. With \(R^{(i)}=0\) and no radial corrections, the Maxwell residuals remain uniformly \(O(\delta^2)\) as \(s\to0\) if and only if \(\partial_a(b/\kappa)=O(\delta^2)\), while cancellation of the first-order scalar residual requires the inverse complex shear to satisfy Eq.~\eqref{c3n:N0scalar-condition}.

By distinguishing the relativistic gradient expansion from the nonrelativistic hydrodynamic expansion and applying the rescaling and frame transformations, we express the leading-order single-copy fields and Weyl curvature in the nonrelativistic \(K\) frame. The resulting leading-order Weyl scalars agree with previous literature~\cite{Keeler2024}, and the algebraic factorizations into two Maxwell spinors have the same structure within the amplitude subfamilies satisfying the corresponding differential constraints.

Several directions merit further investigation. For Petrov type-I fluid geometries, one could examine which pairings of the four distinct principal null directions yield two spinors that each satisfy Maxwell’s equations. Extending the relativistic gradient expansion to second order would allow us to test whether the first-order type-II factorization persists and to determine how the single-copy fields must be corrected. For the fixed-amplitude \(N_0\) branch, an open question is whether radial corrections can cancel the Maxwell and scalar residuals for more general spatial profiles of vorticity and shear while preserving the leading radiative amplitude. Finally, one could seek a joint perturbative solution of the bulk gravitational equations and the Maxwell and scalar field equations, with the Weyl double copy relation satisfied at each retained order.

\acknowledgments
\addcontentsline{toc}{section}{Acknowledgments}

This work is supported by the National Natural Science Foundation of China
(No.~12375059), and the Project of National Astronomical Observatories, Chinese Academy of Sciences (No.E4TG6601).

\end{document}